\documentclass[a4paper,11pt]{article}
\pdfoutput=1 

\usepackage{jinstpub} 

\title{ Construction and beam-tests of silicon-tungsten prototype modules for the CMS High Granularity Calorimeter for HL-LHC}

\author[a,b]{Thorben Quast}

\affiliation[a]{CERN,\\Geneva, Switzerland}
\affiliation[b]{Rheinisch Westf\"alische Technische Hochschule,\\Aachen, Germany}

\emailAdd{thorben.quast@cern.ch}

\abstract{As part of its HL-LHC upgrade program, CMS is developing a High Granularity Calorimeter (HGCAL) to replace the existing endcap calorimeters. The HGCAL will be realised as a sampling calorimeter, including an electromagnetic compartment comprising 28 layers of silicon pad detectors with pad areas of 0.5 - 1.0 $\text{cm}^{2}$ interspersed with absorbers.
Prototype modules, based on 6-inch hexagonal silicon pad sensors with 128 channels, have been constructed and include many of the features required for this challenging detector.
In 2016, beam tests of sampling configurations made from these modules have been conducted both at FNAL and at CERN using the Skiroc2 front-end ASIC (designed by the CALICE collaboration for ILC). In 2017, the setup has been extended with CALICE's AHCAL prototype, a scinitillator based sampling calorimeter, and it was further tested in dedicated beam tests at CERN. There, the new Skiroc2-CMS front-end ASIC was used for the first time. We highlight final results from our studies in 2016, including position resolution as well as precision timing-measurements. Furthermore, the extended setup in 2017 is discussed and first results from beam tests with electrons and pions are shown.}

\collaboration[c]{on behalf of the CMS collaboration}

\proceeding{Conference on Calorimetry for the High Energy Frontier\\
	Lyon, France\\
	02 - 06 October 2017}

\begin{document}
	\maketitle
	\flushbottom
	\section{The CMS HGCal upgrade for HL-LHC}	\label{sec:HGCalUpgrade}
	As part of its preparation for the HL-LHC \cite{HLLHC} upgrade, the CMS collaboration \cite{CMSExperiment} is finalising the technical design of its new endcap calorimeter. The replacement of the current calorimeter endcaps is foreseen in 2024-2025 during LHC's long shutdown 3.
	 Driven by the necessity to cope with expected neutron fluences of $10^{16}$ 1 MeV neutron equivalents per $\text{cm}^{2}$  in the $|{\eta}|\approx 3$ region after 3000 $\text{fb}^{-1}$ of integrated luminosity and with the expected pile up environment of up to 200 interactions per bunch crossing, the CMS collaboration has decided for the silicon-based High Granularity Calorimeter (HGCAL) option  \cite{TechnicalReportHGCal}. 
	The HGCAL is designed as a sampling calorimeter with 28 layers of hexagonal silicon sensors  interspersed with lead absorber in the calorimeter endcap electromagnetic compartment.
	This part is followed by a hadronic section using stainless steel as absorber material. Silicon is used in the high-$\eta$ region as active material while the region with lower radiation levels use plastic scintillator tiles with on-tile silicon photomultiplier (SiPM) readout.
	Ultimately, the design aims at excellent quality in the object reconstruction and identification throughout the entire lifetime despite the dense pileup and high radiation environment. \newline
	While the technical design report is under preparation and will contain detailed information on the expected performance and on the technical realisation of this project, this document explains the construction of the silicon-tungsten prototype modules and focusses on their subsequent tests with particle beams in 2016 and 2017. 
	These tests provide a proof of the concept and aim at the validation of Geant4 \cite{GEANT4} simulations of the proposed design with preliminary readout electronics.

	\section{Module construction}
	\label{sec:ModuleConstruction}
	
\begin{figure}[h]
	\centering 
	\includegraphics[width=0.80\textwidth]{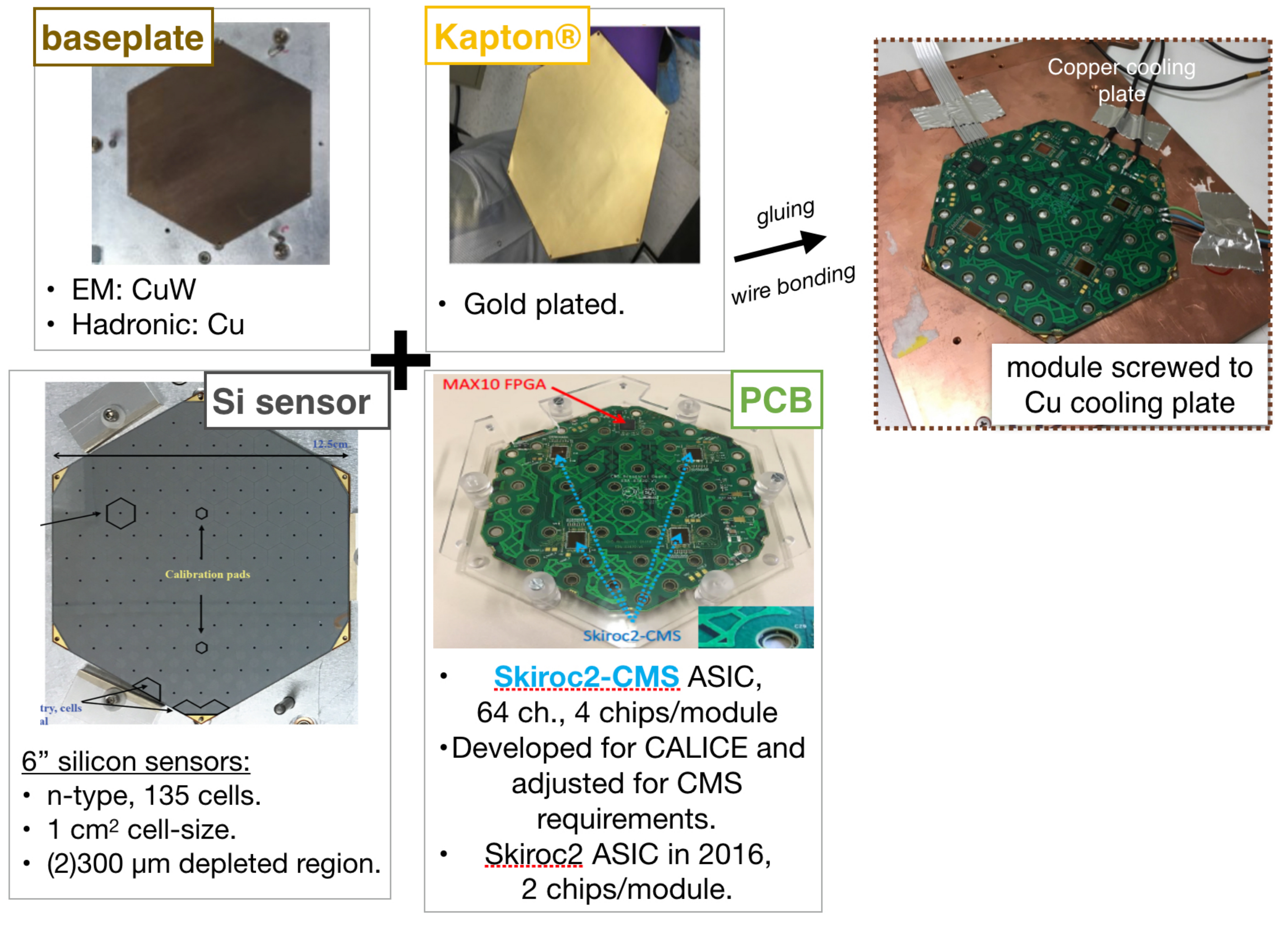}
	\caption{ Overview on the HGCAL beamtest module assembly in 2016/17: Assembly of one module (left), module installation on the cooling plate (top-right).}
	\label{fig:modules}
	\end{figure}

	The sensors used for the beam tests in 2016 and 2017 are made of 6-inch hexagonal silicon n-type wafers. The wafers consist of 135 individual cells. Most of them are full 1$\text{cm}^{2}$-sized  hexagons while cells at the edges are made of alternative shapes. The depletion region of the sensors used for the 2017 (2016) prototype modules is 300 $\upmu$m (200 $\upmu$m) thick. Prior to their assembly into full modules, all sensors have been characterised with respect to their per-cell leakage current and capacitances under different bias voltages. Figure~\ref{fig:modules} illustrates the assembly of one module.
	The module composition starts from copper-tungsten (25\% / 75\%) baseplates in the electromagnetic part while pure copper is used in the hadronic sections. For insulation and biasing purposes, Kapton\textregistered~foil is attached on top before the hexagonal silicon sensors are glued to the baseplates. The foil is plated with gold for biasing the silicon backplane. 
	The readout chips are located on the PCB, which is ultimately wire bonded to the sensitive cells. In the 2017 beam tests, four Skiroc2-CMS ASICs are used per module. Only 32 of the 64 available channels per chip are connected to the sensor. In contrast, the 2016 modules house two Skiroc2 ASICs on an additional PCB, which is plugged into connectors on the first PCB. 
	Finally, the assembled modules are mounted on copper cooling plates and inserted into flexible hanging file structures which include the different absorber plates.

	\section{Summary of beam tests in 2016}
	\label{sec:2016Tests}
	In 2016, a prototype to test the calorimetric performance of the electromagnetic compartment was built and it was tested both at Fermi National Accelerator Lab (FNAL) and CERN. A dedicated summary paper is in preparation and various results have already been reported on, e.g. at \cite{Summary2016}.
	In this section, a reminder of the 2016 setups is given and the two most recent results are highlighted.

	\subsection{Experimental setups at CERN and FNAL}
	\label{subsec:Setup2016}
	The modules tested in 2016 are equipped with the Skiroc2 ASIC \cite{Skiroc2}. As mentioned previously, a hanging file structure was adopted to place these modules flexibly. It allows for the exploration of different sampling configurations. At FNAL the system consisted of 16 sampling layers with a total depth of 15.3 radiation lengths ($\text{X}_{\text{0}}$). This configuration was tested with electrons in the energy range from 4 GeV to 32 GeV. Protons, close to minimum ionizing particles (MIPs), were used to calibrate the system. At CERN, two different setups were investigated. Both of them comprised eight sampling layers. The first configuration had the modules placed between 6 $\text{X}_{\text{0}}$ and 15 $\text{X}_{\text{0}}$, whereas the second one covered a longitudinal depth from 5 $\text{X}_{\text{0}}$ to 27 $\text{X}_{\text{0}}$. 
	Furthermore, a series of tests to demonstrate the instrinsic timing capabilities of the silicon sensors were performed at FNAL and at CERN. We tested irradiated diodes of varying thicknesses, along with unirradiated full sensors.. A dedicated fast digitizer was installed to retrieve the pulse shape arising from signals in the silicon due to incident electrons on upstream absorbers.

	\subsection{Example results}	
	One of the main goals of the 2016 campaign was the validation of the detector simulation using the data taken with the prototypes. Overall, an agreement with the simulated results using the "FTFP\_BERT\_EMM" Geant4 physics list at the percent level is found in terms of electron induced shower shapes, depths and energy resolutions. The two most recent results are presented.
	
	\label{subsec:Results2016}

\begin{figure}[htbp]
		\centering 
		\includegraphics[width=.35\textwidth]{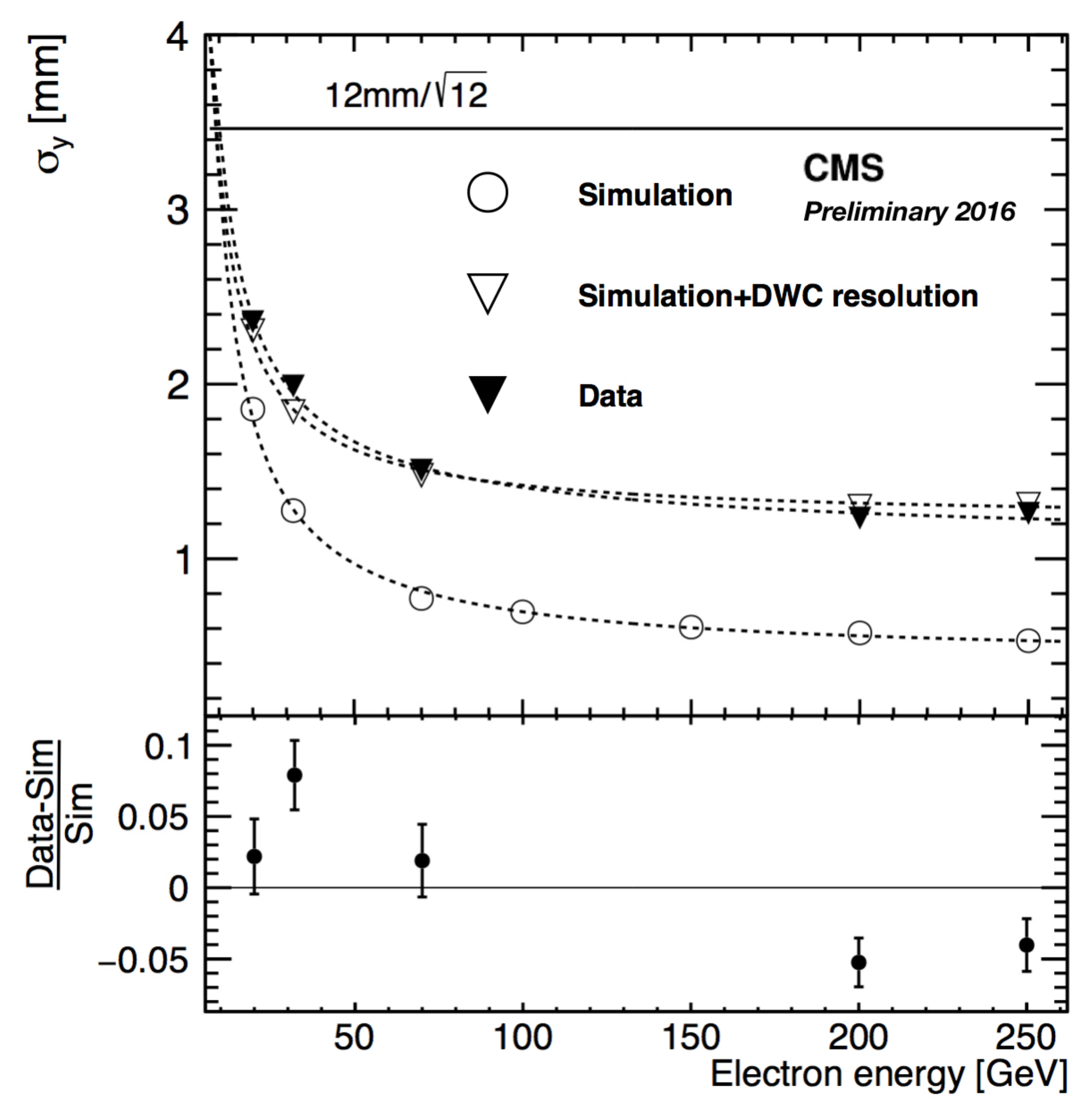}
		\qquad
		\includegraphics[width=.50\textwidth]{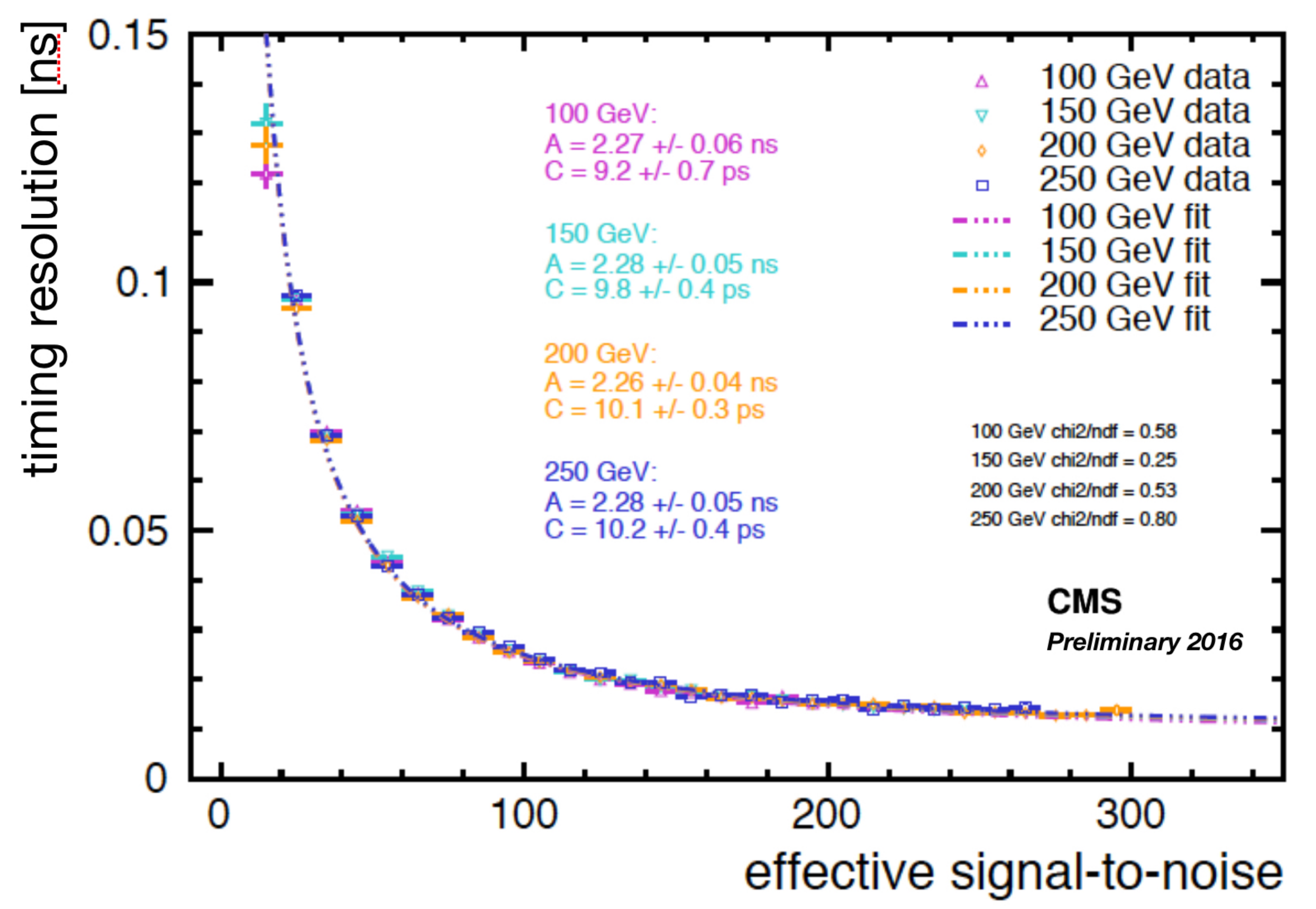}
		\caption{ Left: Spatial precision as a function of electron energy at 6$\text{X}_{\text{0}}$  depth. Right: Intrinsic timing resolution for time differences within a cluster for different electron energies as a function of $\left(S/N\right)_{\text{eff}}$.}
		\label{fig:2016results}
	\end{figure}

	\textbf{Position resolution:}\newline
	For each shower, a logarithmic energy-weighted position barycentre was calculated and compared with the expected incidence position as extrapolated from the upstream delay wire chambers (DWCs) \cite{H2DWCs}. The measured residual width of the y- coordinate as a function of the electron energy after 6 $\text{X}_{\text{0}}$ is shown in the left panel of figure~\ref{fig:2016results}. The resolution of the reference measurement is included in the simulation resulting in an agreement with the recorded data within 5\%. It demonstrates that the intrinsic spatial precision at this depth is around 0.6mm for  electrons above 200 GeV.\newline

	\textbf{Intrinsic timing resolution:}\newline
	The right panel of figure~\ref{fig:2016results} summarises the intrinsic timing precision of an unirradiated full sensor. As an estimate of that precision, the time difference between pairs of neighbouring cells on the sensor was computed for different incident electron energies. The timing precision is displayed as a function of the effective signal-to-noise ratio (S/N), defined as follows:
	\begin{equation}
	 	\label{eq:SN}
	 	\left(S/N\right)_{\text{eff}}=\frac{\left(S/N\right)_{\text{ref}}\left(S/N\right)_{\text{n}}}{\sqrt{\left(S/N\right)_{\text{ref}}^2+\left(S/N\right)_{\text{n}}^2}} 
	 \end{equation}
	Here, the subscript n denotes the sensitive cell and ref denotes the reference measurement.
	It is shown that the intrinsic timing resolution does not depend signficantly on the incident particle's energy at a given signal-to-noise ratio. Both for full hexagonal sensors and for diodes, the resolution is better than 20 ps for $\left(S/N\right)_{\text{eff}}$ > 100, corresponding to 30 MIPs.

	\section{Testbeam campaign in 2017}
	\label{sec:2017Tests}
	In 2017 the setup has been extended by a hadronic section. As planned for the hadronic part of the HGCAL, this section consisted of steel as absorber. Hadronic showers were explored for the first time with a prototype of the HGCAL system. The tested modules were equipped with the Skiroc2-CMS ASIC \cite{Skiroc2CMS}. This chip is an evolution of the Skiroc2, adjusted to suit better the CMS requirements, e.g. with a 5 ns - 80 ns shaping time. The new readout chip allows for the investigation of the timing capabilities of this detector. As one consequence, the data acquisition system had to be revised. Dedicated readout boards have been designed and fabricated. Multiple modules were connected to one readout board through HDMI cables. A dedicated synchronisation board provided a common clock to the readout chips and distributed the trigger signal. The DAQ software is written within the EUDAQ framework \cite{EUDAQ} and makes use of the IPbus control system \cite{IPbus}. Data were taken at CERN's SPS in July and October 2017.

	\subsection{Experimental setup at CERN}
	\label{subsec:Setup2017}
\begin{figure}[h]
	\centering 
	\includegraphics[width=.75\textwidth]{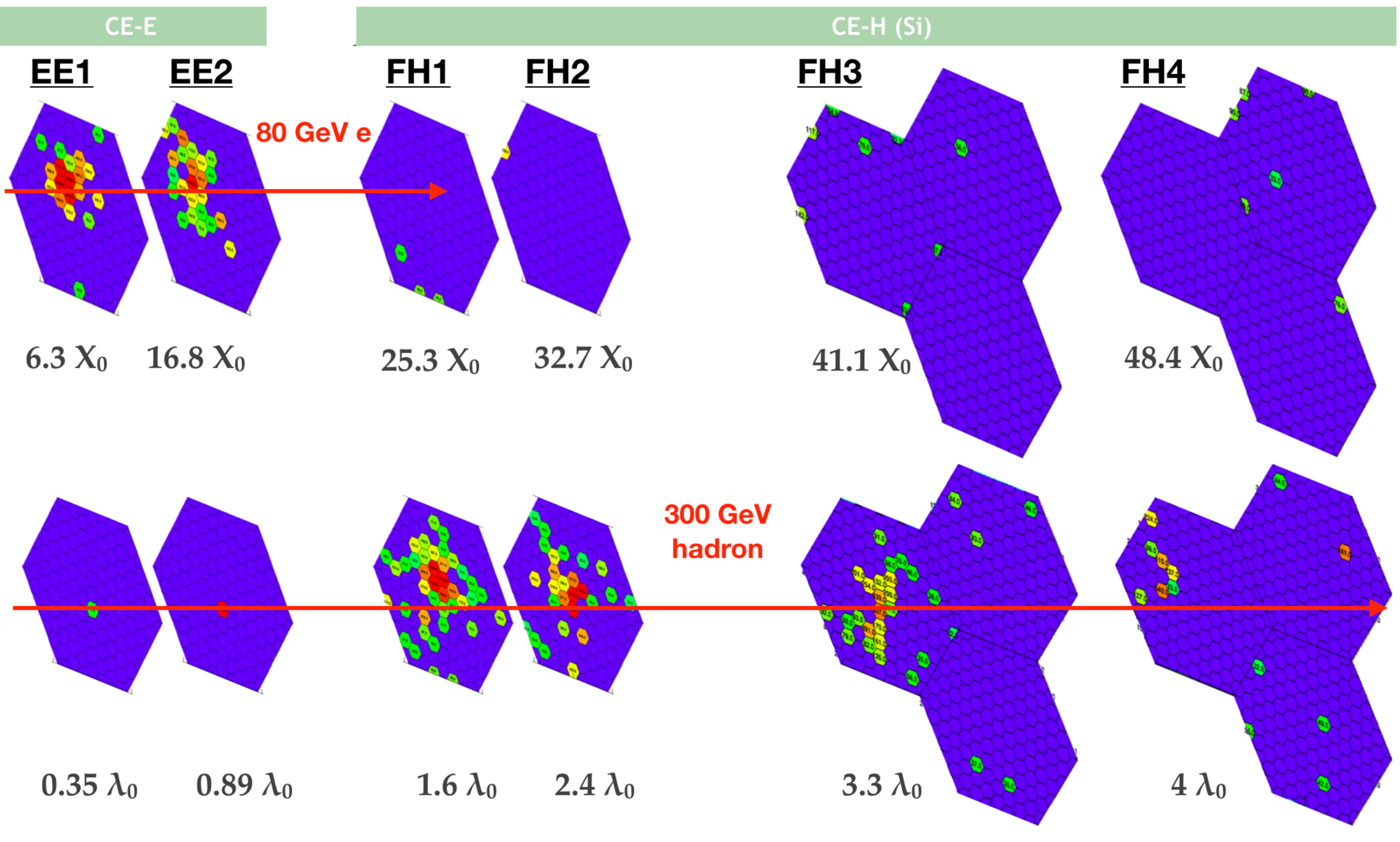}
	\caption{\label{fig:i} Event display for the silicon-based prototype for 80 GeV electrons (top) and 300 GeV hadrons (bottom) for the July 2017 beam test. The AHCAL is not shown.}
	\label{fig:eventdisplay}
	\end{figure}

	The 2017 prototype consisted of parts that represent the silicon-based electromagnetic and hadronic sections of the HGAL design. Due to the lower number of fully assembled, functional modules, only a few sampling layers could be realised for the main test in July.  In order to mimick the scintillator-based hadronic part, the 5 interaction length ($\Lambda_{\text{0}}$) version of CALICE's AHCAL prototype \cite{AHCAL} was placed downstream of the silicon modules and was read out in parallel. The full system was made of two layers in the electromagnetic and four layers in the silicon-based hadronic compartment. The AHCAL itself is made of 12 active layers.	
	The electromagnetic part was 17 $\text{X}_{\text{0}}$ in depth, the silcon based hadronic ended at 4 $\Lambda_{\text{0}}$ and the AHCAL added another 5 $\Lambda_{\text{0}}$. In the July test at the SPS H2 beam line, 80 and 200 GeV electrons as well as hadrons of 100 - 300 GeV were used to probe the system. Dedicated runs with muons were recorded to calibrate the setup with MIP signals.
	Figure~\ref{fig:eventdisplay} illustrates the recorded signals in the silicon-based compartments for 80 GeV electrons, respectively for 300 GeV hadrons.
	Sampling configurations made of 17 and 20 modules could be tested with beam in October. The analysis of these data is ongoing.

	\subsection{First results}	
	\label{subsec:FirstResults2017}

	A high level of common mode noise was present within the 2017 modules. Its origin is still under investigation. Most of this common mode noise can be determined and subtracted on an event-by-event basis using the offline signal reconstruction procedure.  Amplitude spectra for many cells caused by minimum ionising particles were computed, and MIP signals were observed. A feature induced by the common mode noise in these spectra is a structure at around 20 ADC counts, which is made of noisy events that pass the selection criteria prior to the signal pulse reconstruction. 
	\begin{figure}[htbp]
		\centering 
		\includegraphics[width=.40\textwidth]{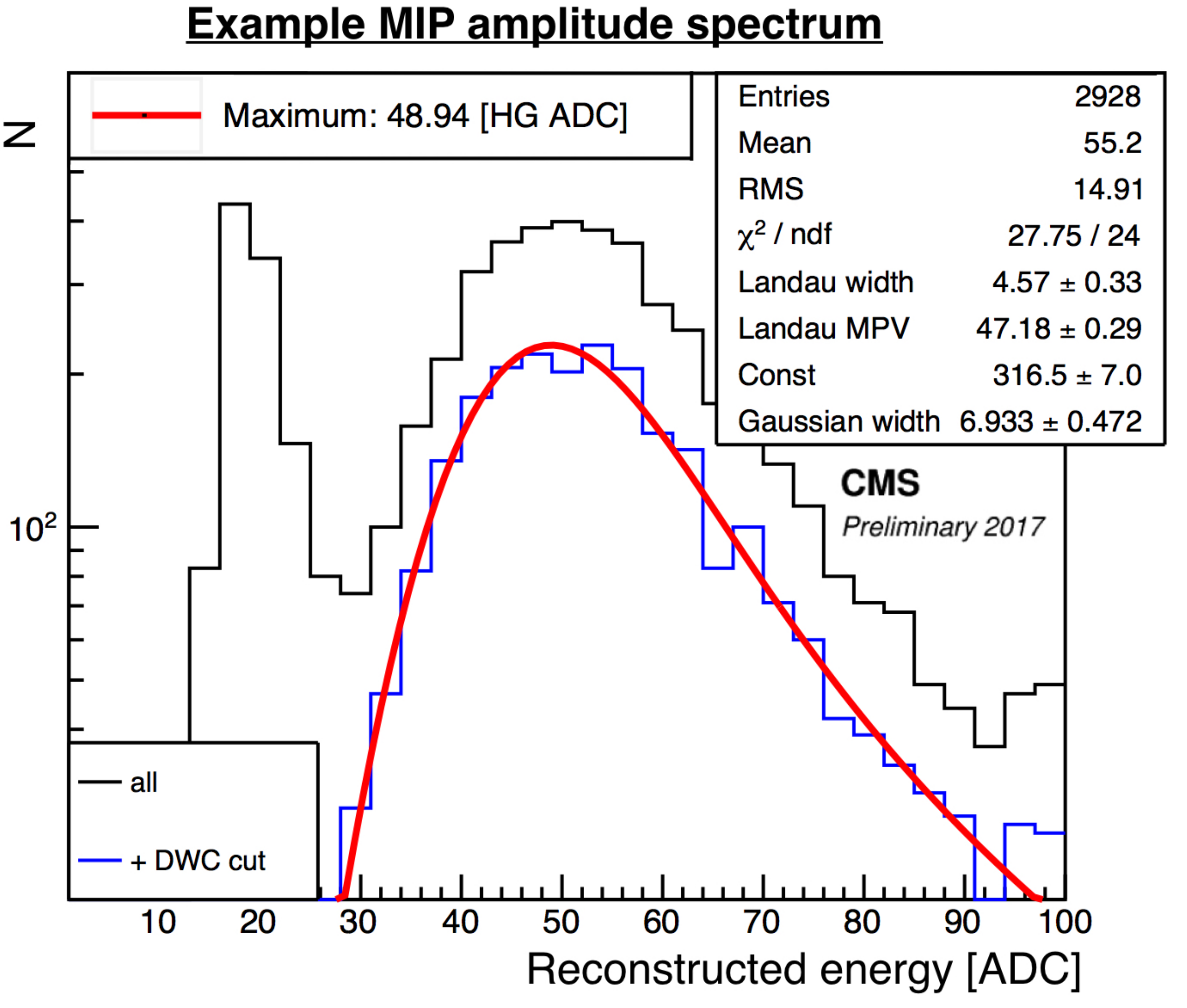}
		\qquad
		\includegraphics[width=.45\textwidth]{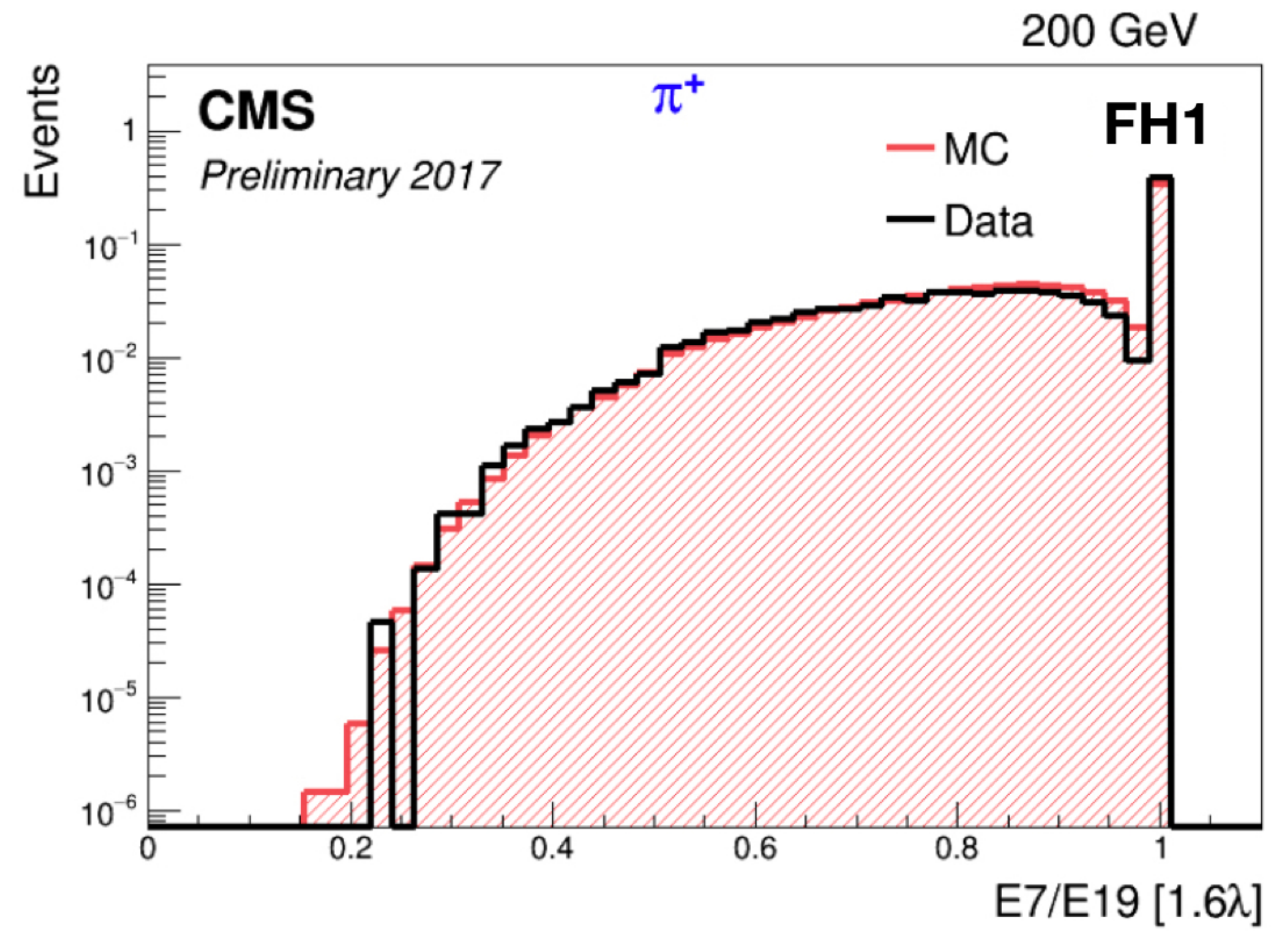}
		\caption{ Left: Amplitude spectra for one cell induced by 150 GeV muons. Right: Transverse shower shape for 200GeV pions in the first layer of the hadronic section.}
		\label{fig:2017results}
	\end{figure}
	Noise in these spectra is removed by selecting only events, for which the channel of interest is hit by the incident particle as extrapolated from four delay wire chambers (DWCs) in the beam line. MIP calibration factors are defined as the maxima of Landau convoluted Gaussians fitted to these spectra as shown in the left panel of figure \ref{fig:2017results}. \newline
	Furthermore, first comparisons of the test beam data to the simulated samples using the \newline FTFP\_BERT\_EMM Geant4 physics list were carried out. For instance, to compare the transverse shower spread, ratios of the energies in central cells and their neighbours are computed. The right panel of figure~\ref{fig:2017results} displays the distribution of E7/E19 for 200 GeV pions at 1.6 $\lambda_{\text{0}}$ depth. Here, E7 is the sum of energy in the most energetic cell  plus the ones from  the surrounding ring of six cells while E19 includes two rings. A good agreement can be seen.

	\section{Outlook}
	\label{sec:Outlook}
	The analysis of the data taken in 2017 and the comparison of the results to the detector simulation is ongoing. In particular, comparisons to different physics lists and the exploration of the timing capabilities with the recorded Skiroc2-CMS ASIC are targeted. After a successful calibration, the energy reconstruction with and without combination of the AHCAL data remains to be studied.
	Further beam tests in 2018 are planned at Fermilab, DESY, IHEP Beijing and CERN. The main objective will be to extend the system with more modules and to further explore the timing performance. Laboratory tests of the modules will investigate the origin of the observed noise and will help to understand the sensors calibration with test pulse injection.


\begin{thebibliography}{99}
		
		\bibitem{HLLHC}
		G. Apollinari, et al., \href{https://cds.cern.ch/record/2116337/files/CERN-2015-005.pdf}{High-Luminosity Large Hadron Collider}, \emph{Technical Design Report CERN-2015-005.} (December 2015) 
		
		\bibitem{CMSExperiment}
		The CMS Collaboration, \href{http://iopscience.iop.org/article/10.1088/1748-0221/3/08/S08004/meta}{The CMS Experiment at the CERN LHC}, \emph{Journal of Instrumentation (Vol 3).}  (August 2008) 
		
		\bibitem{TechnicalReportHGCal}
		D. Contardo, et al., \href{https://cds.cern.ch/record/2020886/files/LHCC-P-008.pdf}{Technical Proposal for the Phase-II Upgrade of the CMS Detector}, \emph{Technical Report CERN-LHCC-2015-010. LHCC-P-008. CMS-TDR-15-02.}  (June 2015)
	
		\bibitem{GEANT4}
		GEANT4 Collaboration, \href{http://inspirehep.net/record/593382}{GEANT4: A Simulation Toolkit}, \emph{Nuclear Instrumentation Methods. A506. p 250-303.}  DOI: 10.1016/S0168-9002(03)01368-8.  (August 2002) 

		\bibitem{Summary2016}
		S. Jain, \href{http://iopscience.iop.org/article/10.1088/1748-0221/12/03/C03011}{Construction and first beam-tests of silicon- tungsten prototype modules for the CMS High Granularity Calorimeter for HL-LHC}, \emph{Journal of Instrumentation (Vol 12).}  (March 2017) 

		\bibitem{Skiroc2}	
		S. Callier, et al., \href{http://iopscience.iop.org/article/10.1088/1748-0221/6/12/C12040/meta}{SKIROC2, front end chip designed to readout the Electromagnetic CALorimeter at the ILC}, \emph{Journal of Instrumentation (Vol 6).}  (December 2011) 	

		\bibitem{H2DWCs}	
		J. Spanggaard, \href{http://cds.cern.ch/record/702443/files/sl-note-98-023.pdf}{Delay Wire Chambers - A Users Guide}, \emph{SL-Note-98-023.}  (1998)	

		\bibitem{Skiroc2CMS}	
		J. Borg, et al., L. Raux, T. Sculac and D. Thienpont, \href{http://iopscience.iop.org/article/10.1088/1748-0221/12/02/C02019/meta}{SKIROC2\_CMS an ASIC for testing CMS HGCAL}, \emph{Journal of Instrumentation (Vol 12).} (February 2017) 	

		\bibitem{AHCAL}
		C. Graf for the CALICE Collaboration, \href{https://arxiv.org/pdf/1711.03796.pdf}{Performance of a Highly Granular Scintillator-SiPM Based Hadron Calorimeter Prototype in Strong Magnetic Fields}. (November 2017) 

		\bibitem{EUDAQ}
		E. Corring, \href{https://www.eudet.org/e26/e28/e86887/e86890/EUDET-Memo-2010-01.pdf}{EUDAQ Software User Manual}, \emph{EUDET-Memo-2010-01.}  (2010)

		\bibitem{IPbus}	
		C. Ghabrous Larrea, et al., \href{http://iopscience.iop.org/article/10.1088/1748-0221/10/02/C02019}{IPbus: a flexible Ethernet-based control system for xTCA hardware}, \emph{Journal of Instrumentation (Vol 10).}  (February 2015) 	
			
		
		
		
	\end{thebibliography}
\end{document}